\documentclass[showpacs,amsmath,amssymb,twocolumn,prl,floatfix]{revtex4}
\usepackage[dvips]{graphicx}
\input{epsf}

\begin{document}

\def\tr#1{\mathrm{Tr}\left(#1\right)}

\title{Chaotic dynamics of a Bose-Einstein condensate coupled to a qubit}

\author{J. Martin$^{1,2,3}$, B. Georgeot$^{1,2}$ and D. L. Shepelyansky$^{1,2}$}

\affiliation{\mbox{$^{1}$Universit\'e de Toulouse, UPS, Laboratoire de
Physique Th\'eorique (IRSAMC), F-31062 Toulouse, France} \\
\mbox{$^{2}$CNRS, LPT (IRSAMC), F-31062 Toulouse, France} \\
\mbox{$^{3}$Institut de Physique Nucl\'eaire, Atomique et de
Spectroscopie, Universit\'e de Li\`ege, 4000 Li\`ege, Belgium}}

\date{December 30, 2008}

\begin{abstract}
We study numerically the coupling between a qubit and a
Bose-Einstein condensate (BEC) moving in a kicked optical lattice,
using Gross-Pitaevskii equation. In the regime where the BEC size is
smaller than the lattice period, the chaotic dynamics of the BEC is
effectively controlled by the qubit state. The feedback effects of
the nonlinear chaotic BEC dynamics preserve the coherence and purity
of the qubit in the regime of strong BEC nonlinearity.  This gives
an example of an exponentially sensitive  control over a macroscopic
state by internal qubit states. At weak nonlinearity quantum chaos
leads to rapid dynamical decoherence of the qubit. The realization
of such coupled systems is within reach of current experimental
techniques.
\end{abstract}

\pacs{05.45.Mt, 67.85.Hj, 37.10.Jk, 42.50.Dv}
\maketitle

The dynamics of Bose-Einstein condensates (BEC) in the regime of
quantum chaos is now actively investigated by several experimental
groups in the world \cite{phillips,summy,auckland,sadgrove}.  These
experiments implement the quantum version of the Chirikov standard
map (kicked rotator) \cite{chirikov} by means of kicked optical
lattices. This model exhibits interesting phenomena such as
dynamical localization of diffusive energy growth, which has been
first observed in cold atoms~\cite{raizen}. However, for cold atoms
the dynamics is described by the linear Schr\"odinger equation,
while for BEC nonlinear effects are present that can be studied in a
wide range of situations with Gross-Pitaevskii equation
(GPE)~\cite{pitaev}. Due to this nonlinearity the dynamics of BEC
can become truly chaotic with appearance of exponential instability
\cite{pikovsky,martin}, provided the size of BEC is smaller than the
optical lattice period. In many cases, a correct description of the
system is obtained by accounting for the internal structure of the
atoms forming the BEC. Such BEC must be seen as quantum objects with
several components. In the situation where there are two components,
the system becomes equivalent to a BEC coupled to a qubit. This
leads to the fundamental problem of the dynamics of a truly chaotic
BEC coupled to a qubit and controlled by it. At the same time, this
dynamics generates a feedback on the purity and evolution of the
qubit, which may significantly affect its properties. Since the BEC
dynamics is exponentially unstable, it has the potential to become a
detector with exponential sensitivity to the qubit state. The spinor
states of BEC are now actively studied experimentally (see
e.g.~\cite{cornell,bigelow}), such that in principle it would allow
to realize the coupling between a qubit and chaotic BEC dynamics.
The decoherence of a qubit coupled to a BEC has also been discussed
in the context of solid state physics devices~\cite{gurvitz}.

To describe the dynamics of a BEC coupled to a qubit in a kicked
optical lattice we use GPE \cite{pitaev}
\begin{align}\label{GP}
    i\hbar\frac{\partial}{\partial t}\left(
    \begin{array}{c}
    \psi_{1} \\
    \psi_{0} \\
    \end{array}
    \right)
    ={}\Big(-\frac{\hbar^2}{2m}\frac{\partial^2}{\partial x^2}+(k+\varepsilon\,\sigma_z)\cos (\kappa x) \; \delta_T(t)\nonumber\\
   +\delta\,\sigma_x\Big)
    \left(\begin{array}{c}
    \psi_{1} \\
    \psi_{0} \\
    \end{array}\right)-\left(\begin{array}{c}
    (g_{11}|\psi_{1}|^2+g_{10}|\psi_{0}|^2)\psi_{1} \\
    (g_{01}|\psi_{1}|^2+g_{00}|\psi_{0}|^2)\psi_{0} \\
    \end{array}\right)
\end{align}
The first term on the r.h.s.~accounts for the free evolution of the
cold atoms, the second one represents the effect of the optical
lattice of period $\lambda =2\pi/\kappa$, $\delta_T(t)$ being a
periodic delta-function with period $T$, and the third one
corresponds to the driving of the qubit with a strength $\delta$.
The last term describes the nonlinear interactions between atoms
treated in the framework of GPE for spinor BEC. We start
considerations with the case $g_{11}=g_{10}=g_{01}=g_{00}=g$, where
the nonlinear parameter $g=Ng_{1D}$ is determined by the number $N$
of atoms in the condensate and the effective 1D coupling constant
$g_{1D}=-2a_0\hbar\omega_{\perp}$, with $\omega_{\perp}$ the radial
trap frequency and $a_0$ the 3D scattering length. The spinor
wavefunction is normalized to one ($\int_{\mathbb{R}}
(|\psi_0(x)|^2+|\psi_1(x)|^2)dx=1$). In the following, we express
the momentum of atoms in recoil units and time $t$ in units of the
kick period $T$ by setting $\hbar=m=\kappa=1$ in Eq.~(\ref{GP}). In
absence of atom-atom interactions at $g=0$, Eq.~(\ref{GP}) reduces
to the usual kicked rotator model with classical chaos parameter
$K=kT$ and effective Planck constant $T$ (see
e.g.~\cite{chirikov,martin}). The first three terms on the
r.h.s.~correspond to the linear evolution of the kicked rotator
coupled to a qubit.  The strength of the kick depends on the
orientation of the qubit due for example to different detunings
between the atomic transition frequencies and the laser frequency
\cite{raizen}. This linear model was studied in \cite{averin} where
it was shown that quantum chaos in the kicked rotator leads to rapid
decoherence of the qubit. Here we study the effects of a
nonlinearity $g$ induced by interactions between atoms in the BEC.

We consider an initial state of the form
$|\Psi\rangle=|\psi\rangle\otimes|\phi\rangle$ where
$|\phi\rangle=\alpha |0\rangle + \beta |1\rangle$ is the internal
state of the BEC and $|\psi\rangle$ its motional state given by the
soliton distribution
\begin{equation}\label{soliton}
    \psi(x,t)=\frac{\sqrt{g_0}}{2}
\frac{\exp\left(ip_0(x-x_0-p_0t/2)+ig_0^2t/8\right)}
{\cosh\left(g_0(x-x_0-p_0t)/2\right)}
\end{equation}
taken at time $t=0$.  For $k=0$ and  $g_0=g$ this is the exact
solution of Eq.~(\ref{GP}), which describes the propagation of a
soliton with constant velocity $p_0$ \cite{refsol}. Numerical
simulations are performed using the second order SABA
integrator~\cite{Laskar} for which the error scales as
$\mathcal{O}(\Delta t^3)$ with time step $\Delta t$. Computations
have been done with $\Delta p=1/64$, total number of states
$N_S=2^{15}-2^{16}$ and time step $\Delta t\approx 0.01/(g+1)$.

\begin{figure}
\begin{center}
\includegraphics[width=.95\linewidth]{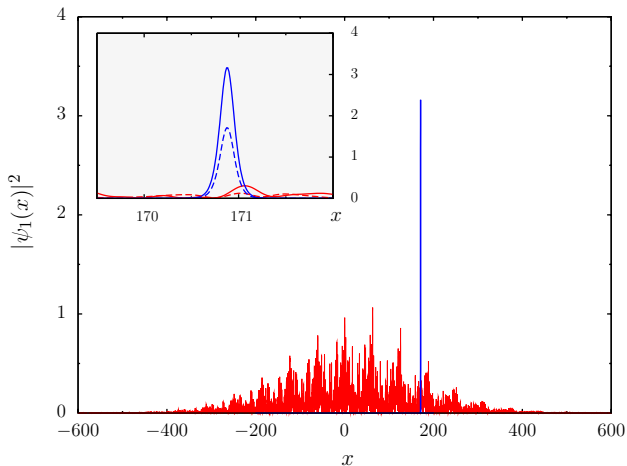}
\end{center}
\caption{(Color online) Probability density $|\psi_{1}(x)|^2$ at
$t=10$ for $k=3.4$, $\varepsilon=0.5$, $T=2$, $\delta=0.2$ showing a
broad distribution at
 $g=0$ (red/grey curve, magnified by a factor 100) and a BEC soliton at $g=20$
(blue/black curve). The initial state is given by
Eq.~(\ref{soliton}) at $t=0$  for $g_0=20$, $x_0=0.1$ and $p_0=0.05$
and with qubit state $|\phi\rangle=(|0\rangle+|1\rangle)/\sqrt{2}$.
Inset shows data on a smaller scale near the soliton center, with
full curves for $|\psi_{1}(x)|^2$ and dashed curves for
$|\psi_{0}(x)|^2$. } \label{fig1}
\end{figure}

\begin{figure}
\begin{center}
\includegraphics[width=.95\linewidth]{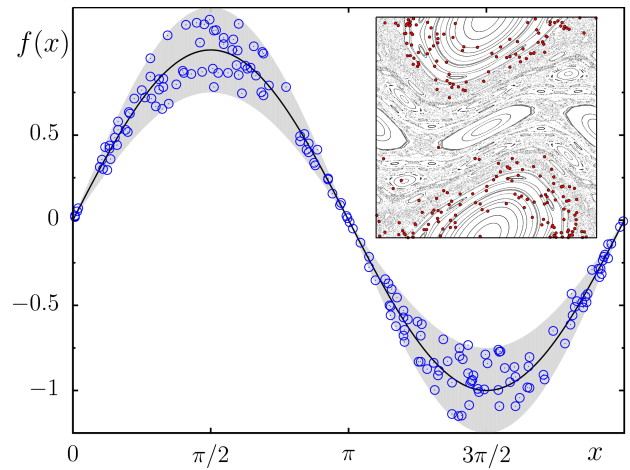}
\end{center}
\caption{(Color online) Circles show the kick function $f(x)$
obtained from the BEC dynamics of Eq.~(\ref{GP}) with parameters
$k=1$, $\varepsilon=0.25$, $T=4$, $\delta=0.2$ and $g=20$. Initial
state is the same as in  Fig.~\ref{fig1}. The black solid line shows
$f(x)=\sin x$ while the grey area, induced by the qubit, corresponds
to the region delimited by the curves $(1\pm\varepsilon/k)\sin x$.
Inset: Poincar\'e section of the standard map at $K=1$ (small dots),
BEC positions are shown by large red/grey dots for parameters $k=1$,
$\varepsilon=0.5$, $T=1$, $\delta=0.05$ and $g=20$ and initial state
as in the main Figure.
} \label{fig2}
\end{figure}

At moderate values of $k$, a narrow distribution with a solitonic
form (\ref{soliton}) spreads rapidly in space and momentum in
absence of interactions between atoms ($g=0$).  In contrast, for
sufficiently large $g$, the shape of the soliton is only slightly
perturbed even after a large number of kicks (see Fig.~\ref{fig1}).
In this regime, the center $(x_c,p_c)$ of the BEC follows the
dynamics described by Chirikov standard map \cite{pikovsky}:
$\bar{p}_c=p_c + k\sin x_c\;;\; \bar{x}_c=x_c+\bar{p}_c T$, where
bars denote the values of the soliton position and velocity after
one kick iteration. Accordingly, the soliton dynamics becomes
chaotic for $K=kT \gtrsim 1$. In presence of the qubit dynamics
imposed by the Rabi splitting $\delta$, the kick function
$f(x_{t})=(x_{t+1}-2x_{t}+x_{t-1})/(kT)$ starts to depend on the
internal qubit state (here $x_t$ is the average position of all
atoms from the two components at time $t$). This leads to a
splitting of $f$ amplitude between the two values
$(1\pm\varepsilon/k)\sin x$ determined by the qubit state.  In
presence of Rabi splitting $\delta$, the amplitude $f$ varies
between these two limiting values due to nontrivial qubit dynamics
as is clearly seen in Fig.~\ref{fig2}. Due to the qubit
oscillations, the soliton can even penetrate inside classically
integrable regions (see inset of Fig.~\ref{fig2}). Inside the
chaotic region a soliton trajectory in the phase space is
exponentially sensitive to the qubit internal quantum state.


\begin{figure}
\begin{center}
\includegraphics[width=.95\linewidth]{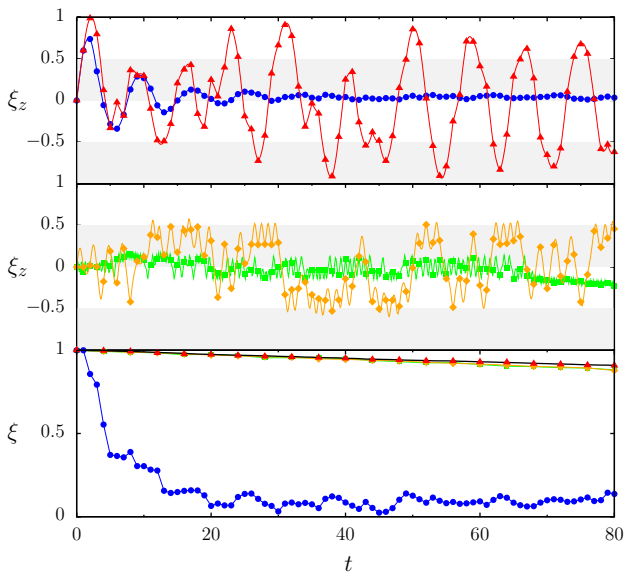}
\end{center}
\vglue -0.50cm\caption{(Color online) Qubit polarization $\xi_z$
(top and middle panels) and purity $\xi$ (bottom panel) as a
function of time for $k=3.4$, $\varepsilon=0.5$, $T=2$, and
$\delta=0.2$. Data are shown for: (top panel)
$g_{11}=g_{10}=g_{01}=g_{00}=g$ with $g=20$ (red triangles) and
$g=0$ (blue circles); (middle panel) $g_{00}=g_{11}=21$ and
$g_{01}=g_{10}=19$ (orange diamonds) and $g_{00}=g_{11}=30$ and
$g_{01}=g_{10}=10$ (green squares). In the bottom panel, the black
solid line shows the solitonic part $W_s$ (see text), other symbols
are as in top and middle panels. Initial state is the same as in
Fig.~\ref{fig1}. } \label{fig3}
\end{figure}

The GPE nonlinearity $g$ also influences strongly the qubit
dynamics.  Indeed, for $g=0$ the purity of the qubit
$\xi(t)=|\tr{\rho(t) \vec{\sigma}}|$ decays rapidly to almost zero
due to quantum chaos in the dynamics of atoms, as can be seen in
Fig.~\ref{fig3}.  The decay is linked to the Lyapunov exponent of
classical chaotic dynamics as discussed in \cite{averin}.  In
contrast, for large values of $g$, the purity of the qubit decays
very slowly, remaining close to $1$ after almost one hundred kicks
(see Fig.~\ref{fig3}).  In this regime, the purity decay is
determined by the decay of the solitonic part of the BEC. Indeed,
the latter can be defined as $W_s=\int_{x_c-10/g}^{x_c+10/g}
(|\psi_0(x)|^2+|\psi_1(x)|^2)dx$ with $x_c$ taken as the soliton
center position, and the data in Fig.~\ref{fig3} show that it decays
in the same way as the purity $\xi$.  Since the lifetime of the
soliton $t_s$ can be rather large (predicted to be $t_s \sim
g^4/k^2$ in \cite{pikovsky}), this implies that the qubit can remain
pure for a very long time.

\begin{figure}
\begin{center}
\includegraphics[width=.95\linewidth]{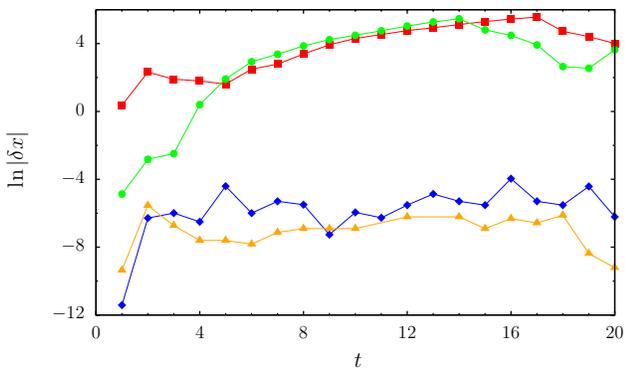}
\end{center}
\caption{(Color online) Distance $\delta x$ between solitons with
the qubit in the states up $\psi_1(x,t)$ and down  $\psi_0(x,t)$ as
a function of time $t$ for $g_{00}=g_{11}=g=40$, $g_{01}=g_{10}=0$
(red squares), $g_{00}=g_{11}=g=39.5$, $g_{01}=g_{10}=0.5$ (green
circles), $g_{00}=g_{11}=g=28$, $g_{01}=g_{10}=12$ (blue diamonds),
and $g_{00}=g_{11}=g_{01}=g_{10}=g=20$ (orange triangles).
Other parameters are as in main part of Fig.~\ref{fig1}. }
\label{fig4}
\end{figure}

In the regime where the qubit purity is high, the probability
oscillates between the two internal states of the BEC.  This leads
to oscillations of the polarization of the qubit $\xi_z =\tr{\rho
\sigma_z}$, which are clearly seen in Fig.~\ref{fig3}. These
oscillations remain stable with respect to small variations of
coupling matrix elements $g_{ij}$ (e.g.\ $g_{10}=g_{01}=19<
g_{00}=g_{11}=g=21$ in Fig.~\ref{fig3}). However for a significant
asymmetry $g_{10}=g_{01}=10, g_{00}=g_{11}=g=30$ the nonlinear shift
strongly modifies the unperturbed Rabi frequency and the amplitude
of oscillations of $\xi_z(t)$ becomes strongly suppressed, even if
the purity remains close to unity. In this regime the distance
between the centers of the two solitons with the qubit in  up and
down states remains small compared to their size (see
Fig.~\ref{fig4}). Only for small values of $g_{ij}=g$ the purity
starts to drop rapidly and the oscillations of $\xi_z(t)$ decay
quickly due to dynamical decoherence induced by quantum chaos. When
the coupling matrix has small off-diagonal elements
(e.g.~$g_{01}=g_{10} \leq 0.5$, see Fig.~\ref{fig4}),
the solitons for the two components
move independently from each other and the purity drops to zero as
soon as the solitons are separated.

\begin{figure}
\begin{center}
\includegraphics[width=.95\linewidth]{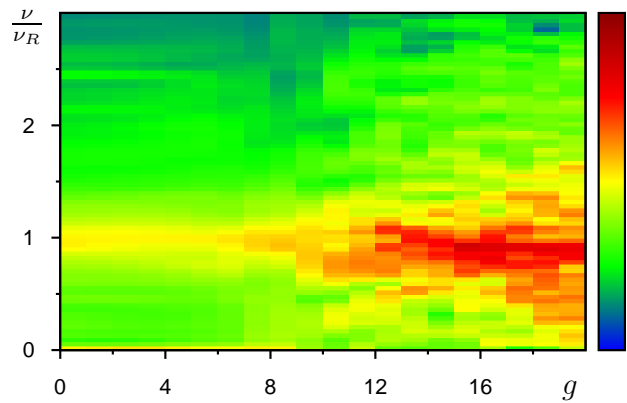}
\end{center}
\caption{(Color online) Density plot of the spectral density
$S(\nu)$ of $\xi_z(t)$ expressed in arbitrary units and shown as a
function of the frequency $\nu$ and the nonlinear parameter $g$ for
$k=3.4$, $\varepsilon=0.5$, $T=2$, $\delta=0.2$, and the rescaled
Rabi frequency $\nu_R=T \delta/\pi$. Initial state is the same as in
Fig.~\ref{fig1}.
} \label{fig5}
\end{figure}

To better analyze the properties of these qubit oscillations, we
determined the variation of spectral density $S(\nu)$ of $\xi_z(t)$
as a function of $g$.  This spectrum is shown in Fig.~\ref{fig5}.
For small values of $g$, the spectral density is very low since
quantum chaos leads to a rapid decay of qubit oscillations and
purity. In contrast, for larger values of $g\geq 10$ a prominent
peak at the Rabi frequency
 $\nu_R=T \delta/\pi$ becomes well pronounced, even if the soliton
dynamics is strongly chaotic ($K=kT=6.8$).

\begin{figure}
\begin{center}
\includegraphics[width=.95\linewidth]{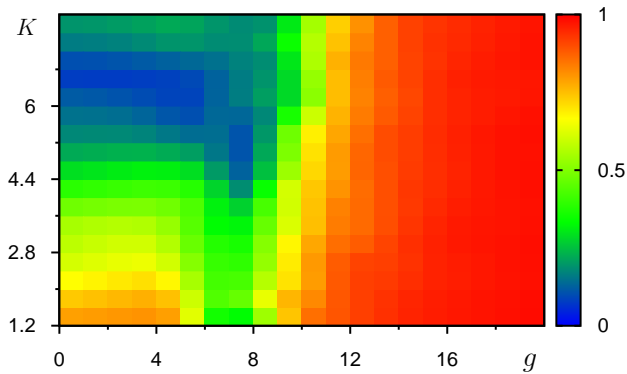}
\end{center}
\caption{(Color online) Density plot of the qubit purity
$\xi$ after $t=20$ kicks as a function of
the chaos parameter $K=kT$ and the nonlinear parameter $g$ for
$T=2$, $\delta=0.2$ and $\varepsilon=0.5$. Initial state is the same
as in  Fig.~\ref{fig1}.} \label{fig6}
\end{figure}

The interplay of quantum chaos and GPE nonlinearity and their effects
on the purity of the qubit are displayed in Fig.~\ref{fig6}.  For small
$g < 10$, the purity is destroyed with the increase of the
chaos strength represented by the parameter $K$.  On the contrary,
for  $g \geq 10$ the purity remains close to $1$ even for large values
of $K$ when the soliton dynamics is strongly chaotic.

This behaviour can be qualitatively understood by the following
arguments. From Eqs.~(\ref{GP})--(\ref{soliton}) the coupling energy
of the single-component soliton at $g_0=g$ is $E_s=-g^2/8$
\cite{landau}.  For $g_{11}\approx g_{10}\approx g_{01}\approx
g_{00} \approx g \gg 1$ the energy of attraction between the two
components is of the same order of magnitude, thus being
significantly larger than the difference $E_k$ between the kinetic
energies induced by the kick term in Eq.~(\ref{GP}). Indeed,
the latter can be estimated as $E_k \sim  p \Delta p \sim
2k\varepsilon\sqrt{t}$; for $k \approx 3, t=20, \varepsilon=0,5$
this gives $E_k \approx 13$ and the condition $E_s\approx E_k$ gives
the critical value of $g_c\approx 10$. This estimate for $g_c$ is
compatible with the numerical results displayed in Fig.~\ref{fig6},
even if this approach is rather simplified. In addition, we should
also note that according to the well-known result \cite{refsol} the
initial state (\ref{soliton}) has a coupled solitonic state for
$g>0.18g_0$. For the conditions of Fig.~\ref{fig6} this gives a
critical $g \approx 3.6$ that is significantly smaller than the
numerical value of $g_c \approx 10$. We attribute this difference to
the internal dynamics of qubit which is not taken into account by
the estimate above. In the case $g_{11}=g_{00}=g \gg g_{01}=g_{10}$
the coupling energy between solitons can be obtained by degenerate
perturbation theory which gives $E_s \sim -g_{01}g$.  In the case
where $E_k > E_s$, the solitons start to separate exponentially fast
due to classical chaotic dynamics.  Such kind of behaviour can be
potentially used to develop detectors with exponential sensitivity
to the qubit state.

In conclusion, our results show that the coupling of a qubit with a
chaotic BEC dynamics leads to rather nontrivial behaviour.  On
one hand, the qubit can control the dynamics of the BEC, moving it
from chaotic to integrable regimes and vice versa. The BEC position
is exponentially sensitive to the state of the qubit. In spite of
chaotic motion of BEC, the purity of the qubit is well-preserved in
this situation. Such an exponential sensitivity of a global BEC
position to an internal quantum state appears due to the absence of
second quantization for BEC motion in GPE. In presence of such
quantization the whole system is described by a linear Schr\"odinger
equation and the chaotic dynamics of BEC can continue only on a
relatively short Ehrenfest time scale as it is discussed in
\cite{martin}. On the other hand, in the regime of weak BEC
nonlinearity, quantum chaos leads to rapid dynamical decoherence of
the qubit. The realization of such coupling between a qubit and a
BEC with chaotic dynamics is within reach of current experimental
techniques.

We thank D.~Gu\'ery-Odelin and T.~Lahaye for useful discussions. We
thank CalMiP for access to their supercomputers and the French ANR
(project INFOSYSQQ) and the EC project EuroSQIP for support. J.M.\
thanks the Belgian F.R.S.-FNRS for financial support.

\end{document}